\documentclass[letterpaper,10pt,conference]{ieeeconf}  % Comment this line out if you need a4paper
\IEEEoverridecommandlockouts                              % This command is only needed if 
\usepackage{cite}
\usepackage{amsmath,amssymb,amsfonts}
\usepackage{times}
\usepackage{latexsym}
\usepackage{array}
\usepackage{longtable}
\usepackage{url}
\usepackage[T1]{fontenc}
\usepackage[utf8]{inputenc}

\usepackage{microtype}
\usepackage{colortbl}
\usepackage{xcolor}
\usepackage{graphicx}
\usepackage{textcomp}
\usepackage{arydshln}
\usepackage{booktabs}
\usepackage{threeparttable}

\usepackage{hyperref}

\usepackage{amssymb}
\usepackage{bbm}
\usepackage{multirow}
\usepackage{graphicx}
\usepackage{booktabs}
\usepackage{stfloats}
\usepackage{xspace}
\usepackage{pifont}
\usepackage{pgfplots}
\pgfplotsset{compat=1.18} 
\usetikzlibrary{decorations.text}
\newtheorem{definition}{Definition}

\newcommand{\ours}[0]{TGAge\xspace}

\title{\ours: A Phenotypic Aging Clock Based on Tabular Foundation Models and the Gompertz Law for Assisted Living Facility Residents}

\author{%Anonymous Authors
Chengzhuo Li$^{1}$, Rongjia Zhou$^{1}$, Shanshan Wang$^{1}$, Runze Yan$^{1}$, Carl Yang$^{1}$, Xiaobo Sun$^{1}$,\\ Huiwen Xu$^{1}$, and Jiaying Lu$^{1}$\textsuperscript{*}
\thanks{$^{1}$Emory University, Atlanta, GA 30322, USA.}
\thanks{\textsuperscript{*}Corresponding Author: Jiaying Lu, Ph.D. (jiaying.lu@emory.edu)}
}

\begin{document}

\maketitle

\begin{abstract}
% 8 pages in total.
Biological age captures overall health status beyond chronological age, yet existing biological aging clocks are difficult to apply to assisted living facility (ALF) residents due to scarce epigenetic data. We propose TGAge, a second-generation phenotypic aging clock that estimates risk-equivalent biological age from routinely collected clinical measurements. TGAge employs a tabular foundation model to predict all-cause mortality risks over discrete time-to-death intervals and converts the resulting mortality-risk ratio into biological age acceleration via the Gompertz law. We develop and evaluate TGAge on a large cohort of ALF residents curated from a proprietary, multi-state long-term care database. TGAge outperforms  clinical indices and established biological aging clocks in survival prediction. Moreover, TGAge-derived age acceleration is significantly associated with multiple aging-related comorbidities. These findings support TGAge as a practical, interpretable, and highly predictive biological aging clock for vulnerable ALF populations using only routinely collected clinical measurements.

\end{abstract}

% \begin{IEEEkeywords}
% phenotypic aging clock, tabular foundation model, Gompertz law, long-term care
% \end{IEEEkeywords}

% Main body
\section{Introduction}

%Para1: long term care facility and ALF residents. The potential utility of age clock model for ALF residents
Population aging is a global problem that poses escalating health, social, and economic challenges. The resulting surge in demand for care has made long-term care facilities an indispensable component of modern health systems~\cite{kaye2010long,lu2025patterns}. Within this landscape, assisted living facilities (ALFs) provide a supportive, home-like environment for individuals who need help with personal care, medication management, or other activities of daily living but do not require continuous skilled nursing~\cite{han2017variation}. In this context, a non‑invasive, accessible, and robust metric that captures an individual’s pace of aging and health deterioration would be of considerable clinical value, particularly for guiding timely interventions.

%Para2: biological age and age acceleration.
A promising basis for such a metric is biological age (BA).
Chronological age (CA), defined as the number of years a person has lived, has been commonly used in clinical practice as a proxy for disease risk, physiological resilience, and treatment suitability. However, individuals of the same chronological age can exhibit markedly different health trajectories and physiological states owing to genetic, environmental, and disease-related factors~\cite{levine13Clock}. BA aims to quantify this heterogeneity by capturing an individual’s biological and functional status, which CA alone often fails to explain.
Previous studies have shown that accelerated BA is associated with worse health outcomes, including physiological deterioration, functional decline, and accumulated cellular damage~\cite{fong2026reframing}. Therefore, accurate estimation of BA holds promise for improving risk stratification, treatment personalization, and long-term care planning in aging populations.

% para3: develop aging clock in ALF population. Gaps and challenges.
In this study, we aim to develop a biological aging clock tailored to assisted living facility (ALF) residents that reflects personalized health risks. Existing biological aging clocks have demonstrated clinical utility in cancer populations~\cite{jiang2025integrating,yan2026predicting}, largely because these populations have richer epigenetic and clinical data available. Many established clocks employ machine learning algorithms to integrate these data into a single measure of biological age~\cite{levine2018epigenetic}. However, applying these clocks to the ALF setting presents distinct challenges. Clinical measurements for ALF residents are often recorded less frequently and with less comprehensive coverage, and epigenetic profiling is generally unavailable because of its prohibitive cost. These limitations underscore the need for a biological aging clock that relies only on routinely collected, accessible clinical measurements while remaining robust to irregular sampling, motivating our proposed approach.

\begin{figure*}[t!]
  \centering
  \includegraphics[width=0.9\textwidth,
  trim={1.8cm 0cm 2cm 2.5cm},
  clip]{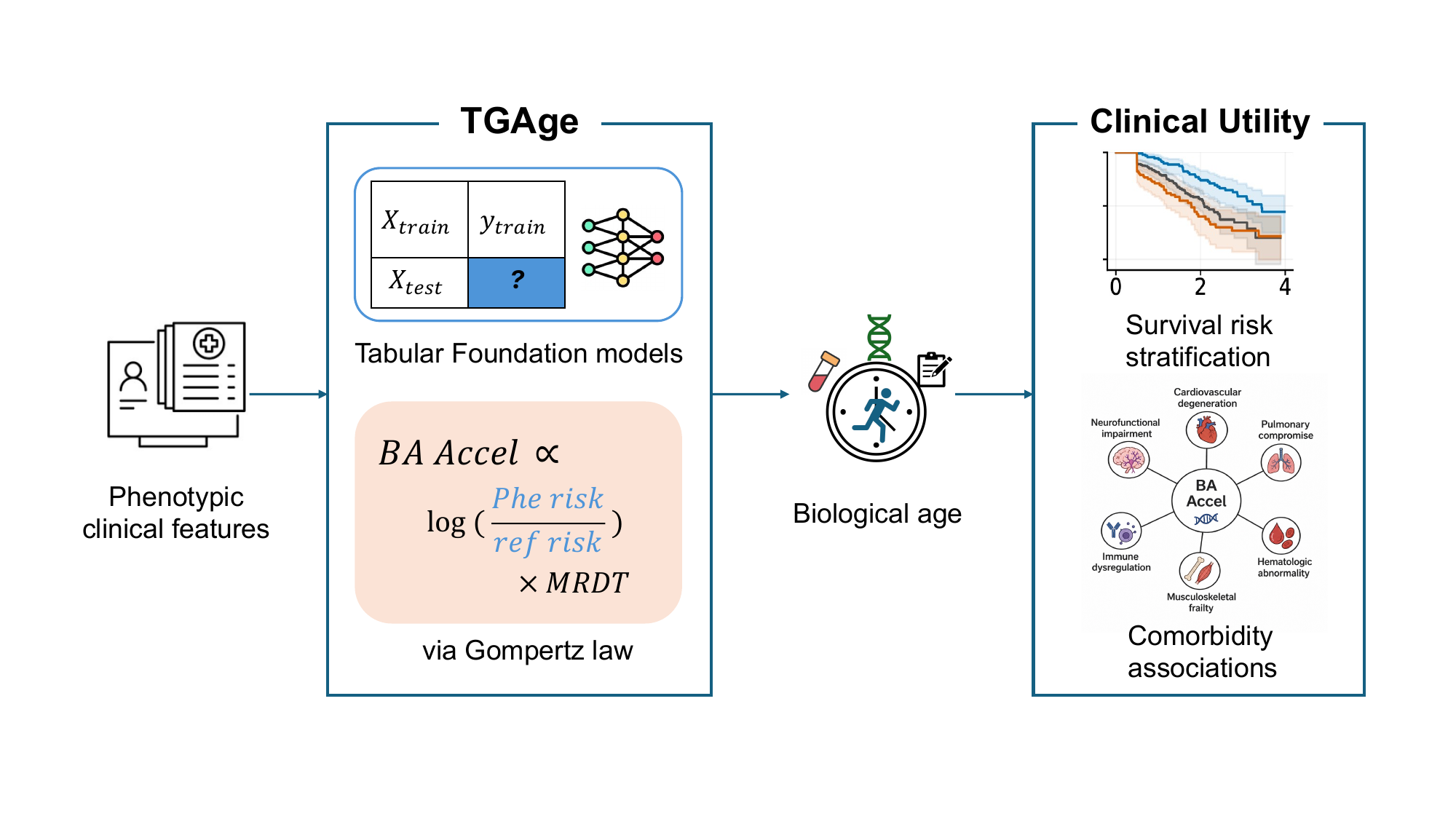}
  \vspace{-1.4cm}
  \caption{An overview of the study design and proposed TGAge clock model.}
  \label{fig:schema}
  \vspace{-0.5cm}
\end{figure*}

%para4: summary our model design
We propose \textbf{\ours} (Tabular Gompertz Age), a second‑generation biological aging clock that uses a tabular foundation model~\cite{hollmann2025tabpfn} to estimate all‑cause mortality risk from clinical phenotypes and then transform that risk into a risk‑equivalent biological age through the Gompertz law~\cite{gompertz1825}. Unlike first‑generation biological aging clocks, which are trained to predict chronological age and inherently conflate biological variation with statistical model error, \ours follows the risk‑equivalent age paradigm: it is explicitly trained on a clinically meaningful endpoint (\textit{i.e.}, mortality) and thus captures the systemic vulnerability that defines aging.
By integrating routinely collected clinical measurements through a foundation model and anchoring the output to Gompertz's mortality rate doubling time, \ours yields an interpretable aging measure that directly reflects an individual’s risk of adverse health outcomes. This makes \ours an actionable indicator of overall physiological status, aligned with the goal of guiding personalized care in aging populations (Fig.~\ref{fig:schema}).

% para5: summary our contirbution
In this study, we curate an ALF resident cohort from a U.S.-based long-term care electronic health record (EHR) database~\cite{pcc_rwd_2025}, consisting of approximately 10,000 subjects for model development and evaluation. The cohort covers ALFs across all 50 states and 2 territories of the U.S., with an average of 165 facility‑living days per encounter. 
We extensively evaluate \ours against widely used clinical indices~\cite{wolff2002prevalence,bynum2005influence}, established first-generation biological aging clocks~\cite{levine13Clock,li2025LLMAge}, and second-generation clocks~\cite{fong2024principal}. Experimental results demonstrate that \ours provides accurate survival prediction and consistently outperforms all baselines. Furthermore, \ours-derived BA acceleration captures clinically meaningful health risks: it effectively stratifies low‑ and high‑risk groups and shows strong associations with aging‑related comorbidities. Finally, ablation studies support the effectiveness of key technical components of \ours.

\section{Related Work}

\subsection{Phenotypic Aging Clock Models}
%Researchers have benchmarked DNA methylation based clock models~\cite{kriukov2025computagebench}.
Phenotypic aging clocks estimate an individual’s BA using clinically accessible phenotypic features. These clocks have evolved across multiple generations and are primarily distinguished by their training targets.
First-generation clocks are trained to directly predict an individual's CA, based on the hypothesis that, when fitted on a large and diverse population, the residual between predicted and actual CA can capture inter-individual variation in biological aging. Various machine learning models have been explored to instantiate first-generation clocks, including linear regression~\cite{hollingsworth1965linearregression}, random forest~\cite{kodera2025randomforest}, support vector regression~\cite{zhang2024svr}, XGBoost~\cite{borsky2025xgb}, neural networks~\cite{huang2025dnn}, large language models~\cite{li2025LLMAge}, etc. 
However, as CA itself is not a direct measure of health status, these models may fail to fully reflect age-related physiological decline. 
This limitation motivated the development of second-generation clocks, which aim to measure biological aging by predicting future health risks rather than merely reconstructing CA. Most second-generation clocks rely on the ``Gompertz law''~\cite{gompertz1825} as a theoretical basis, as it captures the exponential increase in mortality hazard across adult ages. 
For instance, Levine’s phenotypic clock~\cite{levine2018epigenetic} uses 9 clinical measurements and Gompertz proportional hazards models to estimate 10-year mortality risk. 
More recently, PCAge~\cite{fong2024principal} extends this risk-equivalent aging framework by estimating mortality risk from principal components of clinical features and converting the resulting risk ratio into biological age acceleration.

%To address the limitation, researchers have proposed to use surrogate age instead of actual chronological age as the training target. For instance, ~\cite{levine2018epigenetic} use 9 clinical measurements with specified coefficients to derive their surrogate age, and then train another clock model to fit on the derived surrogate age.

%These clocks define true BA as ``Gompertz age'', or the age commensurate with an individual's future risk of dying from all intrinsic causes. 
%Organ specific age clocks.

%\subsection{Clinical Index for Aging Measurement}
%introduce the indices. Especially the ones we will use and compare in this study.

\subsection{Tabular Foundation Models for Survival Analysis}
Survival analysis aims to model time-to-event outcomes, such as time until death, while properly accounting for censored observations. The Cox proportional hazards model~\cite{cox1972regression} remains one of the most widely used survival models due to its interpretability and ability to estimate covariate-associated hazard ratios. However,the standard Cox model relies on the proportional hazards assumption and typically models covariate effects in a log-linear form, which may limit its ability to capture complex nonlinear relationships and high-order feature interactions in large-scale biomedical datasets.
To address these limitations, recent studies have explored machine learning and deep learning approaches for survival analysis, including Random Survival Forests~\cite{kodera2025randomforest}, DeepSurv~\cite{katzman2018deepsurv}, DeepHit~\cite{lee2018deephit}, and transformer-based survival models~\cite{wang2022survtrace}.
More recently, tabular foundation models~\cite{hollmann2025tabpfn,qu2025tabicl} have emerged as a promising paradigm for learning from heterogeneous tabular data with strong generalization ability. Several recent studies have begun adapting these models to survival analysis. 
For example, TabSurv~\cite{vu2025tabsurv} formulates time-to-event prediction as a regression problem over uncensored observations, 
MITRA~\cite{kim2026mitra} casts survival analysis as a series of binary classification tasks over discretized time intervals, and SIC~\cite{seletkov2026sic} combines TabICL~\cite{qu2025tabicl} with a DeepHit survival prediction head. These studies highlight the potential of tabular foundation models for survival prediction, but their role in risk-equivalent BA estimation remains underexplored. This motivates our use of a tabular foundation model to derive Gompertz-law-based BA from mortality risk estimates.

\section{Study Design}
\subsection{Data Source}

% comment from Jun 23: SNF few days -- several weeks; 
% ALF has <18 residents. 2025 (Oct) may not complete. 

We use the proprietary PointClickCare (PCC)~\cite{pcc_rwd_2025} database as the data source for this study. PCC is one of the largest EHR systems serving long-term and post-acute care (LTPAC) facilities across North America. The database contains longitudinal health records for more than 20 million unique patients collected over two decades of routine clinical care, comprehensively covering aging and vulnerable populations in skilled nursing facilities, ALFs, and other LTPAC settings. It includes rich patient-level information on demographics, diagnoses, medication orders, laboratory results, vital signs, functional assessments, cognitive assessments, and activities of daily living, with near real-time updates and longitudinal follow-up. These characteristics make PCC a valuable resource for studying aging trajectories, mortality risk, and BA estimation in large-scale, real-world populations.

\subsection{Cohort Construction}

We choose ALF residents as the target population for this study. ALFs represent a major component of the LTPAC continuum, providing a home-like residential environment that combines housing, personal care, and limited medical assistance while emphasizing residents' independence and quality of life. Compared with residents in skilled nursing facilities or nursing homes, ALF residents generally require less intensive medical care and retain greater autonomy in daily activities, including the ability to select their preferred pharmacy and participate in community-based services. Unlike nursing homes, which employ standardized resident assessments and federally regulated reporting systems, ALFs typically provide individualized services tailored to residents' needs upon admission. Their scope of care and regulatory requirements vary substantially across states, resulting in heterogeneous care practices and documentation standards. Nevertheless, ALFs routinely collect essential demographic, clinical, medication, and functional information to support care planning and service delivery.
ALF residents are generally older adults with diverse health conditions and varying levels of functional decline, making them an important population for studying aging trajectories and mortality risk. At the same time, the availability of routinely collected clinical and medication data enables the development of data-driven models for estimating mortality hazards and BA. Table~\ref{tab:data_source_overview} presents an overview of all ALF residents in the PCC database.

\begin{table}[htbp!]
\centering
\caption{Overview of all ALF residents in the data source.}
\label{tab:data_source_overview}
\vspace{-0.2cm}
\resizebox{\linewidth}{!}{%
\begin{tabular}{lc}
\toprule
\textbf{Characteristic} & \textbf{Value} \\
\midrule
\multicolumn{2}{l}{\textbf{Population}} \\
Number of encounters & 1,236,208 \\
Number of residents & 1,202,978 \\
Age, mean (SD) & 76.3 (17.2) \\
%\cdashline{1-1}
\textit{Sex} \\
\quad Female, n (\%) & 776,975 (64.6\%) \\
\quad Male, n (\%) & 426,003 (35.4\%) \\
%\cdashline{1-1}
\textit{Race} \\
\quad American Indian or Alaska Native, n (\%) & 5,173 (0.43\%) \\
\quad Asian, n (\%) & 16,842 (1.40\%) \\
\quad Black or African American, n (\%) & 127,997 (10.64\%) \\
\quad Hispanic or Latino, n (\%) & 30,436 (2.53\%) \\
\quad Native Hawaiian or Other Pacific Islander, n (\%) & 2,887 (0.24\%) \\
\quad White, n (\%) & 687,381 (57.14\%) \\
\quad Other, n (\%) & 332,262 (27.62\%) \\
%Ethnicity / Hispanic & Data no such info\\ 
\midrule
\multicolumn{2}{l}{\textbf{Facility}} \\
Number of ALFs & 26,618 \\
Number of states and territories & 52 \\
Mean living days (SD) & 164.9 (532.8) \\
%Study period & 2015--2025 \\
\midrule
\multicolumn{2}{l}{\textbf{Survival Outcomes}} \\
Deaths observed (\%) & 281,980 (22.8\%) \\
Mean follow-up years (SD) & 4.1 (3.0) \\
Mean survival time years (SD) & 1.9 (1.7) \\
% Censoring rate (\%) & 77.2\% \\
\bottomrule
\end{tabular}
}
\end{table}

Specifically, the source cohort is defined at the encounter level. We include residents with any qualifying ALF admission date between 2015/01/01 and 2025/12/31. This source cohort comprises 1,236,208 unique encounters corresponding to 1,202,978 unique residents. To develop a phenotypic biological aging clock, we retain encounters in which all required 24 clinical measurements are available, yielding a final analytic sample of 10,516 encounters (0.85\%) from 10,467 residents (0.87\%).
To balance physiological coverage and sample size, the 24 clinical measurements include: 
(1) four \textit{vital signs}: pulse, systolic blood pressure, diastolic blood pressure, and body mass index; (2) nine \textit{hematology} measurements: hematocrit, hemoglobin, red blood cell count, red cell distribution width, mean cell volume, mean cell hemoglobin, mean cell hemoglobin concentration, platelet count, and white blood cell count; and (3) eleven \textit{clinical chemistry} measurements: blood urea nitrogen, creatinine, sodium, potassium, chloride, total calcium, total bilirubin, total protein, aspartate aminotransferase, alkaline phosphatase, and glucose.
We adopt a temporal split to evaluate the generalizability of the clock model: the training set consists of encounters from 2015 to 2021, and the test set consists of encounters from 2022 to 2025. In addition, we retain a reference set for developing the Gompertz law‑based approach. This reference set also spans 2015–2021 and is the superset of the training set before the complete 24‑feature filter. Table~\ref{tab:patient_cohorts} presents the detailed characteristics of each subset.

\begin{table}[htbp!]
\centering
\caption{Characteristics of the selected study cohort.}
\label{tab:patient_cohorts}
\resizebox{\linewidth}{!}{%
\begin{tabular}{lcccc}
\toprule
\textbf{Split} & \textbf{\#Encounters} & \textbf{\#Residents} & \textbf{\#Female (\%)} & \textbf{Age (SD)} \\
\midrule
% below three are before 24-feature fltering
% Overall & 1,236,208  & 1,202,978  & 776,975 (62.9\%)  & 76.3 (17.2)  & -- \\
% Training & 696,837 & 679,011 & 448,224 (64.3\%) & 76.6 (15.2) & -- \\
% Testing & 539,371  & 528,441 & 328,751 (61.0\%) & 75.9 (19.5) & 0.039 \\
%Overall & 10,516 & 10,467 & 7,154 (68.0\%) & 78.2 (13.9) \\
Reference & 696,837 & 679,011 & 448,224 (64.3\%) & 76.6 (15.2)\\
Training & 5,608 & 5,603 & 3,903 (69.6\%) & 77.1 (13.1) \\
Testing & 4,908 & 4,881 & 3,251 (66.2\%) & 79.5 (14.7) \\
\bottomrule
\end{tabular}
}
\vspace{-0.2cm}
% \begin{minipage}{0.95\columnwidth}
% \footnotesize
% \centering
% Age is reported as mean (SD). The 24-feature cohort refers to residents with complete availability of the phenotypic features used for biological age modeling.
% \end{minipage}
\end{table}

\subsection{Notation and Problem Setup}

We formalize the biological age estimation task as follows.

\begin{definition}[BA Estimation]
For the $i$-th ALF resident encounter, the model $f_\theta(\cdot)$ uses chronological age $\mathrm{CA}_i$, sex $sex_i$, and phenotypic features $\mathbf{C}_i$ to estimate BA $\hat{y}_i=f_\theta(\mathrm{CA}_i,sex_i,\mathbf{C}_i)$, where $\hat{y}_i\in\mathbb{R}_{>0}$.
\end{definition}

\begin{definition}[Phenotypic Features]\label{def:feature}
An encounter's phenotypic features are represented by a vector $\mathbf{C}_i=[c_{i,1},c_{i,2},\dots,c_{i,d}]$, where each element $c_{i,j}$ is a tabular clinical measurement (e.g., blood pressure, BMI). 
\end{definition}

\begin{definition}[Survival Outcome]\label{def:outcome}
The survival outcome of the $i$-th encounter is represented as $(T_i,\delta_i)$, where $T_i$ denotes the observed follow-up time from the admission date and $\delta_i$ denotes the all-cause mortality indicator. Specifically, $\delta_i=1$ indicates that death is observed during follow-up, while $\delta_i=0$ indicates not observed.
\end{definition}

\section{Methods}

\subsection{Risk-equivalent BA under the Gompertz Approximation}

\ours estimates biological age as an age-equivalent transformation of mortality risk. Under the Gompertz law~\cite{gompertz1825}, adult mortality hazard increases approximately exponentially with chronological age. Therefore, a subject whose mortality risk is higher than that of an age- and sex-matched reference subject can be interpreted as having a positive BA acceleration.
Let $R_i$ denote the relative mortality risk of subject $i$ compared with the reference risk expected from chronological age and sex alone. Following second-generation BA clocks~\cite{levine2018epigenetic,fong2024principal}, we define risk-equivalent BA acceleration as
\begin{equation}\label{eq:delta_ba}
\Delta\mathrm{BA}_i = \frac{\log R_i}{\log 2} \cdot \mathrm{MRDT}_{\mathrm{sex}_i},
\end{equation}
where $\mathrm{MRDT}_{sex_i}$ denotes the mortality-rate doubling time for sex $\mathrm{sex}_i \in \{\text{female}, \text{male}\}$. The term $\log R_i / \log 2$ is the number of mortality-risk doublings relative to the reference, and multiplying it by $\mathrm{MRDT}_{\mathrm{sex}_i}$ converts this quantity into years. 
To obtain sex‑specific $\mathrm{MRDT}_\mathrm{sex}$ values, we fit a sex‑stratified Cox proportional hazards model~\cite{cox1972regression} with CA as the sole predictor on the reference set:
\begin{equation}
h_\mathrm{sex}(t \mid \text{CA}) = h_{0,\mathrm{sex}}(t) \cdot \exp(\beta_{\text{CA}, \mathrm{sex}} \cdot \text{CA}),
\end{equation}
from which the mortality-rate doubling time is estimated as
\begin{equation}
{\mathrm{MRDT}}_\mathrm{sex} = \frac{\log 2}{{\beta}_{\text{CA}, \mathrm{sex}}}.
\end{equation}
Here, $h_{0,\mathrm{sex}}(t)$ is an unspecified sex-specific baseline hazard function, and $\beta_{\text{CA}, \mathrm{sex}}$ is estimated by partial likelihood without requiring a parametric form for $h_{0,\mathrm{sex}}(t)$.
Finally, biological age is obtained by
\begin{equation}
\mathrm{BA}_i = \mathrm{CA}_i + \Delta\mathrm{BA}_i.
\end{equation}

% \jiaying{to introduce it with reference model.} 
% Sex-specific $\mathrm{MRDT}_{s_i}$ values are estimated from the chronological-age coefficient of the age-only proportional Cox hazard model~\cite{cox1972regression}:
% \begin{equation}
%     \widehat{\mathrm{MRDT}}_s=\frac{\log(2)}{\hat{\beta}_{\mathrm{age},s}}.
% \end{equation}

\subsection{Tabular Foundation Model for Mortality Risk Estimation}

Accurate estimation of the relative mortality risk $R_i$ is critical for \ours. $R_i$ is calculated by
\begin{equation}\label{eq:relative_risk}
R_i=\frac{h_{\mathrm{Phe},i}}{h_{\mathrm{ref},i}},
\end{equation}
where $h_{\mathrm{Phe},i}$ denotes the phenotypic-based mortality risk, $h_{\mathrm{ref},i}$ denotes the reference mortality risk. 
Taking advantage of the superior performance of tabular foundation model~\cite{hollmann2025tabpfn,qu2025tabicl} on tabular data, we leverage a prior-data fitted network implementation TabPFN~\cite{hollmann2025tabpfn} to derive the mortality risks. Unlike conventional deep learning models that make inference directly from learned parameters, TabPFN makes inference by approximating Bayesian in-context learning: 
\begin{equation}
\hat{\mathbf{y}}_i=f(\mathbf{X}^{train},\mathbf{y}^{train},\mathbf{X}_i).
\end{equation}

To utilize TabPFN for time-to-mortality risk prediction needed by Eq.~\eqref{eq:relative_risk}, we follow the discrete-time survival model (also known as classification-based survival analysis) setup~\cite{yu2011mtlr,kvamme2019time} to formulate mortality probabilities over discrete follow-up intervals. Specifically, we partition the follow-up time axis into $K$ discrete intervals $\{\tau_k\}_{k=1}^{K}$, and assign each subject to the interval containing their $T_i$ (introduced in Definition~\ref{def:outcome}). Each discrete interval $\tau_k$ is defined as:
\begin{equation}
\tau_k=
\begin{cases}
(t_{k-1},t_k] \quad 1 \leq k < K,\\
(t_{K-1}, \infty) \quad k = K.
\end{cases}
\end{equation}
where $K \in \mathbb{N}^{+}$ and $t_k$ denotes the pre-specified follow-up time cut points, measured in years since the admission date. $0=t_0<t_1<t_2<\dots < t_{K-1}$ are the boundaries for the discrete intervals. 
Under this discrete-time setup, the survival outcome can be transformed into a supervised multiclass classification problem, with $\mathbf{y}_i=(y_{i,1},y_{i,2}\dots,y_{i,K})$ where
\begin{equation}
    y_{i,k}\triangleq \mathbbm{1}(T_i\in \tau_k).
\end{equation}
We implement two multiclass TabPFN risk estimators for Eq.~\ref{eq:relative_risk}. The phenotypic mortality risk model uses all chronological age, sex, and clinical measurements as input features, therefore 
\begin{equation}
\begin{split}
\hat{\mathbf{y}}_{\mathrm{Phe},i}=f(\mathbf{C}_i^{train}\oplus \mathrm{CA}_i^{train}\oplus \mathrm{sex}_i^{train}, \mathbf{y}^{train},\\
    \mathbf{C}_i\oplus \mathrm{CA}_i\oplus \mathrm{sex}_i).
    \end{split}
\end{equation}
Meanwhile, the reference mortality risk model uses only chronological age and sex, 
\begin{equation}
    \hat{\mathbf{y}}_{\mathrm{ref},i}=f(\mathrm{CA}_i^{train}\oplus \mathrm{sex}_i^{train},\mathbf{y}^{train},\mathrm{CA}_i\oplus \mathrm{sex}_i).
\end{equation}
Since both $\hat{\mathbf{y}}_{\mathrm{Phe},i}$ and $\hat{\mathbf{y}}_{\mathrm{ref},i}$ are $K$-dimensional vectors, we need to aggregate them into scalars for Eq.~\ref{eq:relative_risk}. We adopt cumulative summation of interval-specific mortality probabilities as the aggregation operator $\sigma(\cdot)$,
\begin{equation}\label{eq:cum}
    h_{\mathrm{\cdot},i}=\sigma(\mathbf{\hat{y}_{\mathrm{\cdot},i}})=\sum_{k=1}^{K-1} \hat{\mathbf{y}}_{\mathrm{\cdot},i, k}.
\end{equation}
This applies to both $h_{\mathrm{Phe},i}$ and $h_{\mathrm{ref},i}$. The mortality probability of the last discrete interval $\tau_K=(t_{K-1},\infty)$ represents far-future mortality risk that lies beyond typical observation window. Thus, the risk-equivalent biological age acceleration (Eq.~\ref{eq:delta_ba}) is instantiated as
\begin{equation}
    \Delta\mathrm{BA}_i=\frac{\log\frac{\sigma(\mathbf{\hat{y}_{\mathrm{Phe},i}})}{\sigma(\mathbf{\hat{y}_{\mathrm{ref},i}})}}{\log 2}\cdot MRDT_{\mathrm{sex}_i}
\end{equation}
To support our choice of $cumulative(\cdot)$ as the aggregation operator $\sigma(\cdot)$, we conduct ablation study of other possible aggregation operators including $\min(\cdot)$ and $\max(\cdot)$ (see Sec.~\ref{subsec:ablation}).

\section{Experimental Results}

In this section, we evaluate the proposed \ours model on the following research questions (RQ). \textbf{RQ1}: How accurately can \ours predict survival time compared with baseline models? \textbf{RQ2}: What is the clinical utility of \ours-derived age acceleration? \textbf{RQ3}: What insights do ablation studies reveal?

%we examine whether TGAge improves mortality-risk prediction compared with clinical indices, first-generation biological age clocks, and second-generation biological age clock baselines. 
%Second, we assess whether the resulting biological age and age acceleration capture clinically meaningful variation beyond chronological age. 
%Third, we conduct ablation analyses to understand the contribution of different survival-modeling and TabPFN-based design choices. 

\subsection{Implementation Details}
% first introduce baselines; finally introduce exp setup.
\noindent \textbf{The proposed model}. \ours is implemented using TabPFN-V3 with the default model configuration (\textit{e.g.}, 8 estimators).  We implement the discrete time partition using four intervals: $\{(0, 1\textrm{ year}], (1\textrm{ year}, 2\textrm{ year}], (2\textrm{ year}, 3\textrm{ year}], (3\textrm{ year}, \infty)\}$ to best fit our data. 
%The multiclass label uses four mortality-time bins: event-free through 3 years, death within 1 year, death between 1 and 2 years, and death between 2 and 3 years. Subjects censored before 3 years are excluded from classifier fitting because their 3-year time-bin label is unobserved, but they remain in downstream survival evaluation when a score is available. 
Numeric clinical features are z-score standardized within the training set. All predicted mortality probabilities for each time interval are clipped to $[10^{-6},1-10^{-6}]$ before the aggregation operation. 
The anonymized source code for the core \ours implementation is publicly available at \href{https://anonymous.4open.science/r/TGAge}{anonymous.4open.science/r/TGAge}.
% Training-set \ours scores are generated using fivefold out-of-fold prediction; test-set scores are generated by refitting the full and reference classifiers on all eligible labeled training rows.

\noindent \textbf{Compared baselines}. We compare \ours against three baseline groups. 
(1)\textbf{\textit{clinical indices}} that serve as practical, low‑cost indicators of patients' overall health status, including chronological age, the comorbidity index~\cite{wolff2002prevalence}, and the healthcare use index~\cite{bynum2005influence}. The comorbidity index is implemented as a log-plus-one transformation of the number of distinct conditions (i.e., $\log(1+n_{\mathrm{dx}})$), and the healthcare use index as a log-plus-one transformation of number of distinct days with laboratory testing and number of internal transfers within one encounter.
(2) \textbf{\textit{first-generation biological aging clocks}} that are trained to reconstruct chronological age from phenotypic features using various machine learning methods, including Hollingsworth's linear regression~\cite{hollingsworth1965linearregression}, Kodera's random forest~\cite{kodera2025randomforest}, Zhang's support vector regression~\cite{zhang2024svr}, Borsky's XGBoost~\cite{borsky2025xgb}, Huang's neural networks~\cite{huang2025dnn}(which we implement using TabNN), and LMAge~\cite{li2025LLMAge}(which we implement using ClinicalBERT~\cite{alsentzer2019ClinicalBERT}).
(3) \textbf{\textit{second-generation biological aging clocks}} that are optimized for risk-equivalent BA, including PCAge and LinAge~\cite{fong2024principal}. We implement two variants of PCAge, using 10 principal components (PCAge-10PCs) and 20 principal components (PCAge-20PCs).

\noindent \textbf{Evaluation setup}. All experimental results are reported on the test set of the selected ALF cohort described in Table~\ref{tab:patient_cohorts}. Models are trained on the training set; second-generation BA clocks additionally utilize the reference set for model development. It is worth noting that the clinical indices do not require training.
For survival analysis-based evaluation (\textit{e.g.}, RQ1, RQ3), prediction outcomes of all baselines and \ours (\textit{e.g.}, clinical index, BA) are input as covariates into sex-stratified Cox proportional hazards models fitted on the training set. The fitted Cox models are then applied to the test set to evaluate the clock models' performance.
We use Harrell's concordance index (C-index)~\cite{harrell1982cindex} and time-dependent AUC at 1, 2, and 3 years~\cite{uno2007timeAUC} as the primary evaluation metrics. Time-dependent AUC is computed by defining cases as deaths within each horizon and controls as encounters that remain event-free beyond that horizon; inverse-probability-of-censoring weighting is applied, and an administrative censoring date of December 31, 2025 is used to determine horizon eligibility. All metrics are reported separately for females and males.

\begin{table*}[htbp!]
\centering
\caption{Survival prediction performance of \ours and baseline models on the held-out test set.}
\label{tab:exp_survival}
\begin{tabular}{lcccccccc}
\toprule
\multirow{2}{*}{\textbf{Model}} & \multicolumn{4}{c}{\textbf{Female}}  & \multicolumn{4}{c}{\textbf{Male}} \\
\cmidrule(lr){2-5} \cmidrule(lr){6-9}
& \textbf{C-index} & \textbf{1yr AUC} & \textbf{2yr AUC} & \textbf{3yr AUC}
& \textbf{C-index} & \textbf{1yr AUC} & \textbf{2yr AUC} & \textbf{3yr AUC} \\
\midrule
\textit{(Clinical Indices)} \\
Chronological Age & 0.628 & 0.656 & 0.643 & 0.619 & 0.677 & 0.666 & 0.712 & 0.669 \\
Comorbidity Index & 0.634 & 0.668 & 0.639 & 0.620 & 0.674 & 0.666 & 0.704 & 0.666 \\
%Self health index; comments: a questionaire index from NHANES \\
Healthcare Use Index & 0.651 & 0.707 & 0.665 & 0.620 & 0.693 & 0.700 & 0.735 & 0.676 \\
\hline
\textit{(1st-Gen Clocks)}\\
Hollingsworth    & 0.597 & 0.619 & 0.595 & 0.584 & 0.677 & 0.677 & 0.689 & 0.678 \\
Kodera    & 0.565 & 0.560 & 0.553 & 0.566 & 0.625 & 0.600 & 0.630 & 0.614 \\
Zhang   & 0.584 & 0.594 & 0.578 & 0.583 & 0.638 & 0.626 & 0.650 & 0.638 \\
Borsky   & 0.589 & 0.612 & 0.575 & 0.582 & 0.621 & 0.586 & 0.636 & 0.653 \\
Huang & 0.614 & 0.654 & 0.613 & 0.590 & 0.687 & 0.704 & 0.692 & 0.674 \\
LMAge          & 0.561 & 0.581 & 0.560 & 0.552 & 0.616 & 0.608 & 0.612 & 0.615 \\
\hline
\textit{(2nd-Gen Clocks)}\\
PCAge-10PCs & 0.639 & 0.677 & 0.653 & 0.611 & 0.694 & 0.693 & 0.723 & 0.688 \\
% we have 10pc; 20pc; 23/22pc.
PCAge-20PCs & 0.640 & 0.665 & 0.648 & 0.634 & 0.718 & 0.728 & 0.741 & 0.685 \\
LinAge & 0.644 & 0.674 & 0.653 & 0.629 & 0.699 & 0.699 & 0.723 & 0.678 \\
\ours & \textbf{0.683} & \textbf{0.738} & \textbf{0.712} & \textbf{0.639} & \textbf{0.728} & \textbf{0.741} & \textbf{0.756} & \textbf{0.726} \\
\bottomrule
\end{tabular}
\end{table*}

\subsection{\ours Accurately Predicts Survival Time (RQ1)}

To answer RQ1, we evaluate each method using the Cox-based survival framework described above. 
As shown in Table~\ref{tab:exp_survival}, \ours achieves the strongest overall survival-discrimination performance among all compared methods in both female and male ALF residents. In females, the strongest competitor is the healthcare-use index, whereas in males it is the PCAge-20PCs model. Against these strongest sex-specific baselines, \ours improves the average survival-discrimination score across C-index and the three time-dependent AUC horizons by 4.9\% in females and 2.8\% in males. The improvement is particularly notable because the baseline set includes both practical clinical indices and established biological-age clocks, including second-generation risk-equivalent clocks. 
Moreover, \ours shows consistent gains across across the 1-, 2-, and 3-year horizons rather than at a single endpoint. These results support the effectiveness and robustness of the technical design of \ours. 

\begin{figure}[htbp!]
  \centering
  \includegraphics[width=\linewidth]{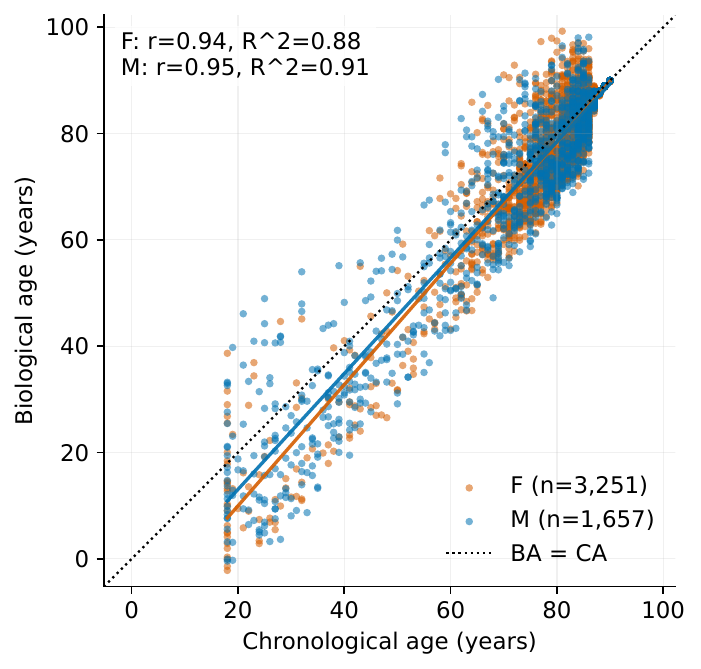}
  \vspace{-0.8cm}
  \caption{Biological age versus chronological age in the held-out test set.}
  \label{fig:ba-vs-ca-test}
\end{figure}

\begin{figure}[htbp!]
  \centering
  \vspace{-0.2cm}
  \includegraphics[width=\linewidth]{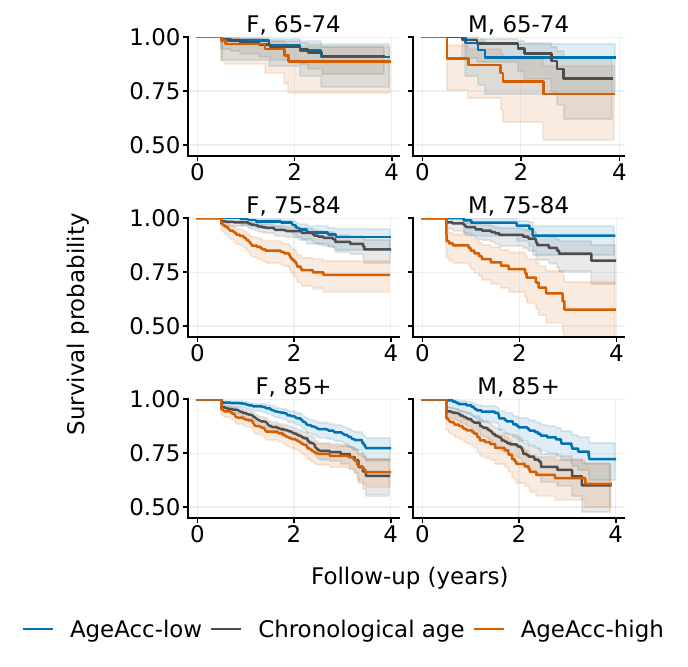}
  \vspace{-0.6cm}
  \caption{Kaplan--Meier survival curves stratified by \ours-derived age acceleration among older adults in the held-out test set.}
  \label{fig:km-ageacc-test}
\end{figure}

%We further examine whether the predicted biological age behaves as a meaningful age-like quantity rather than an arbitrary risk score. 
Next, we evaluate the correlation between CA and \ours-derived BA. A strong but imperfect correlation is desirable: it confirms that the proposed aging clock captures the age‑dependent component of mortality risk, while the residual variation reflects meaningful biological differences.
As shown in Fig.~\ref{fig:ba-vs-ca-test}, \ours-derived BA shows a strong Pearson correlation with CA (average $r=0.94$, average $R^2=0.89$) in the test set. 
Notably, the residual spread (\textit{i.e.}, age acceleration or deceleration) is larger in younger age groups, suggesting that \ours differentiates individuals more strongly when health trajectories are more heterogeneous.

To further assess the prognostic value of \ours, we perform Kaplan--Meier survival analysis~\cite{kaplan1958nonparametric} in older adults ($\mathrm{CA}\geq65$) from the test set. Kaplan--Meier curves provide a non-parametric estimate of survival probability over follow-up while accounting for censored observations, allowing us to visualize whether \ours-derived age acceleration corresponds to distinct empirical survival trajectories. The selected residents are stratified into three groups based on the magnitude of age acceleration: AgeAcc-high (top 25\%), AgeAcc-low (bottom 25\%), and Chronological age (middle 50\%). As shown in Fig.~\ref{fig:km-ageacc-test}, residents in the AgeAcc-high group consistently exhibit worse survival, whereas those in the AgeAcc-low group show more favorable survival. This separation within age-matched strata indicates that \ours does not merely reproduce CA, but captures additional physiological vulnerability directly relevant to mortality risk. 
Taken together, these results answer RQ1 affirmatively: \ours provides a more predictive and clinically meaningful aging-related measurement for ALF residents beyond CA, clinical indices, and existing biological aging clocks.

\subsection{\ours-predicted Age Acceleration Captures Individual Health Risks (RQ2)}

To explore the clinical utility of \ours-predicted BA, we systematically analyze the association between age acceleration and comorbidities. Specifically, we use the \ours cumulative age-acceleration score and test each comorbidity indicator in a separate linear regression model, adjusting for chronological age, sex, and admission date. We then correct the resulting $p$-values using the Benjamini--Hochberg false discovery rate (FDR) procedure~\cite{benjamini1995fdr}.

% explain the findings from Table.
Among 108 evaluated clinical condition categories in our dataset, we identify 14 diagnosis-defined health conditions significantly associated with age acceleration at a strict threshold of FDR $q<0.01$ (Table~\ref{tab:ba_comorbidity_qvalue}). These associations form a multi-system comorbidity signature consistent with aging-related physiological vulnerability. 
Notably, conditions reflecting \textbf{neurofunctional impairment and frailty-related functional loss} show some of the largest effect sizes, including \textit{Paralysis}, \textit{Other acquired deformities}, \textit{Developmental disorders}, \textit{Epilepsy, convulsions}, and \textit{Osteoporosis}; the largest adjusted effect is observed for \textit{Paralysis} ($\beta=4.56$).
\textbf{Cardiovascular aging} is captured by significant positive associations with \textit{Cardiac dysrhythmias}, \textit{Congestive heart failure}, and \textit{Heart valve disorders}. Given their high prevalence in the cohort, these conditions represent clinically important signals.
Markers of \textbf{pulmonary decline, infection, and inflammation} are also prominent: \textit{Respiratory failure, insufficiency, and arrest}, \textit{Pneumonia}, \textit{Bacterial infection}, \textit{Skin and subcutaneous tissue infections}, and \textit{Inflammation and infection of eye} all show significantly positive associations with age acceleration.
Furthermore, conditions such as \textit{Deficiency and other anemia} provide hematologic signals consistent with systemic vulnerability in older adults.
The full spectrum of identified comorbidity conditions aligns closely with known clinical phenotypes of aging, spanning neurofunctional impairment, cardiovascular degeneration, pulmonary compromise, immune dysregulation, hematologic abnormality, and musculoskeletal frailty. The fact that these associations persist even after adjusting for CA indicates that \ours captures aging-related physiological deterioration beyond what can be explained by the passage of time alone. This comorbidity signature provides strong evidence for the biological plausibility and clinical relevance of \ours-derived BA.

\begin{table*}[htbp!]
\centering
\caption{Comorbidities significantly associated with \ours-derived biological age acceleration after FDR correction.}
\label{tab:ba_comorbidity_qvalue}
\begin{threeparttable}
\small
\setlength{\tabcolsep}{4pt}
\renewcommand{\arraystretch}{1.05}

\begin{tabular}{p{0.40\textwidth}rrrrr}
\toprule
\textbf{Comorbidity Condition}
& \textbf{$n$ (\%)}
& \textbf{Adjusted $\beta$}
& \textbf{95\% CI}
& \textbf{$p$-value}
& \textbf{FDR $q$} \\
\midrule
%\multicolumn{6}{l}{\textbf{Associations significant at $q<0.01$}} \\
%\midrule
Paralysis
& 145 (1.4)
& 4.56
& 3.34--5.77
& 1.8e-13
& 9.9e-12 \\

Developmental disorders
& 256 (2.4)
& 2.62
& 1.67--3.56
& 5.7e-8
& 1.0e-6 \\

Other acquired deformities
& 182 (1.7)
& 2.72
& 1.65--3.79
& 6.3e-7
& 8.5e-6 \\

Respiratory failure; insufficiency; arrest
& 244 (2.3)
& 2.34
& 1.41--3.27
& 8.0e-7
& 9.6e-6 \\

Deficiency and other anemia
& 2,010 (19.1)
& 0.85
& 0.50--1.21
& 3.0e-6
& 3.2e-5 \\

Cardiac dysrhythmias
& 2,013 (19.1)
& 0.84
& 0.48--1.20
& 4.8e-6
& 4.8e-5 \\

Pneumonia
& 220 (2.1)
& 2.23
& 1.26--3.21
& 7.3e-6
& 6.6e-5 \\

Epilepsy; convulsions
& 448 (4.3)
& 1.51
& 0.81--2.22
& 2.5e-5
& 1.9e-4 \\

Inflammation; infection of eye
& 118 (1.1)
& 2.85
& 1.52--4.18
& 2.6e-5
& 1.9e-4 \\

Skin and subcutaneous tissue infections
& 216 (2.1)
& 2.09
& 1.11--3.08
& 3.1e-5
& 2.0e-4 \\

Osteoporosis
& 1,146 (10.9)
& 0.97
& 0.51--1.43
& 3.2e-5
& 2.0e-4 \\

Bacterial infection
& 152 (1.4)
& 2.40
& 1.23--3.57
& 5.7e-5
& 3.2e-4 \\

Congestive heart failure (nonhypertensive)
& 1,225 (11.6)
& 0.90
& 0.46--1.34
& 6.0e-5
& 3.2e-4 \\

Heart valve disorders
& 601 (5.7)
& 1.01
& 0.41--1.62
& 0.0010
& 0.0049 \\

\bottomrule
\end{tabular}
\end{threeparttable}
\end{table*}

\begin{table*}[htbp!]
\centering
\caption{Ablation results for two key design choices of \ours.}
\label{tab:ablation_study}
\begin{tabular}{lcccccccc}
\toprule
\multirow{2}{*}{\textbf{Model}} & \multicolumn{4}{c}{\textbf{Female}}  & \multicolumn{4}{c}{\textbf{Male}} \\
\cmidrule(lr){2-5} \cmidrule(lr){6-9}
& \textbf{C-index} & \textbf{1yr AUC} & \textbf{2yr AUC} & \textbf{3yr AUC}
& \textbf{C-index} & \textbf{1yr AUC} & \textbf{2yr AUC} & \textbf{3yr AUC} \\
\midrule
\multicolumn{7}{l}{\textit{(Different Survival Models for Mortality Risk Estimation)}}\\
\ours-TabPFN & \textbf{0.683} & \textbf{0.738} & \textbf{0.712} & \textbf{0.639} & \textbf{0.728} & \textbf{0.741} & \textbf{0.756} & \textbf{0.726} \\
\quad\quad\quad -Cox & 0.644 & 0.674 & 0.653 & 0.629 & 0.699 & 0.699 & 0.723 & 0.678 \\
\quad\quad\quad -DeepSurv & 0.613 & 0.655 & 0.593 & 0.593 & 0.683 & 0.697 & 0.698 & 0.661 \\
\quad\quad\quad -Random Survival Forest & 0.646 & 0.666 & 0.651 & 0.617 & 0.687 & 0.690 & 0.706 & 0.665 \\
\hline
\multicolumn{7}{l}{\textit{(Different Aggregation Operators on Predicted Mortality Risks)}}\\
\ours-Cumulative & \textbf{0.683} & \textbf{0.738} & \textbf{0.712} & \textbf{0.639} & \textbf{0.728} & \textbf{0.741} & \textbf{0.756} & \textbf{0.726} \\
\quad\quad\quad -Max & 0.680 & 0.732 & 0.686 & 0.629 & 0.701 & 0.708 & 0.717 & 0.677 \\
\quad\quad\quad -Min & 0.652 & 0.693 & 0.675 & 0.635 & 0.710 & 0.714 & 0.730 & 0.718 \\
\bottomrule
\end{tabular}
\end{table*}

\subsection{Ablation Analyses (RQ3)}\label{subsec:ablation}

To answer RQ3, we perform two ablation studies targeting two main design choices of \ours: the mortality-risk estimator model and the aggregation operator used to convert the TabPFN-output multiclass mortality-probability vector into a scalar risk estimate. In all ablation experiments, we keep the same model design and experimental settings except for the technical component under evaluation. 
%the same training/testing split, input feature set, full-versus-reference risk-ratio formulation, Gompertz MRDT transformation, and Cox-based survival evaluation framework as in the main experiment.

First, we evaluate whether the choice of mortality-risk estimator model contributes to the performance of \ours. In this ablation, we replace only the survival model used to estimate the phenotypic and reference risks. Specifically, TabPFN is compared against linear Cox proportional hazards model~\cite{cox1972regression} (precisely the same implementation as LinAge), DeepSurv (a popular implementation of neural Cox models)~\cite{katzman2018deepsurv}, and Random Survival Forest~\cite{ishwaran2008random}. 
As shown in Table~\ref{tab:ablation_study}, the TabPFN-based \ours achieves the best observed performance across all metrics. These findings highlight that our discrete-time adaptation of the tabular foundation model for risk estimation is a meaningful design choice with clear technical contribution. 
%These results indicate that the tabular foundation model is not merely an interchangeable implementation choice, but contributes substantially to the mortality-risk estimation component of \ours.
%For female residents, TGAge-TabPFN obtained a C-index of 0.683, compared with 0.644 for the Cox-based variant, 0.613 for DeepSurv, and 0.646 for random survival forest. For male residents, TGAge-TabPFN achieved a C-index of 0.728, compared with 0.699, 0.683, and 0.687 for the corresponding alternatives. The advantage was also consistent across time-dependent AUCs, particularly at the 1-year and 2-year horizons. These results indicate that the tabular foundation model is not merely an interchangeable implementation choice, but contributes substantially to the mortality-risk estimation component of TGAge.

The second ablation focuses on the aggregation operator used to summarize the TabPFN-output multiclass mortality-probability vector into a scalar risk estimate required by the downstream \ours pipeline. We compare three aggregation operators: (1) cumulative aggregation (Eq.~\ref{eq:cum}), which takes the sum of the predicted probabilities over all in-horizon death intervals; (2) max aggregation, which uses the largest predicted mortality probability among all intervals; and (3) min aggregation, which uses the smallest predicted mortality probability among all intervals.
Cumulative aggregation achieves the highest observed performance among the aggregation operators (Table~\ref{tab:ablation_study}).
It is worth noting that cumulative aggregation also has a clear survival interpretation: it estimates the cumulative mortality risk at a chosen prediction horizon (in this case, 3-year).

\section{Conclusion}

In this study, we introduce \ours, a second‑generation biological aging clock that combines a tabular foundation model with a Gompertz‑based risk-to-age transformation to estimate risk‑equivalent age from routine clinical data. \ours provides an interpretable, data‑driven assessment of overall health beyond chronological age. It consistently outperforms established aging clocks and clinical indices, and ablation analyses confirm the contribution of its key design choices. The resulting age acceleration also demonstrates clear clinical utility through risk stratification and significant associations with comorbidities. Importantly, \ours requires only routinely collected EHR variables, making it practical for resource‑limited long‑term care settings. Its foundation‑model backbone further positions it to benefit from future advances in tabular deep learning.

Several limitations should be acknowledged. First, \ours has so far been evaluated only among assisted living facility residents; we thus plan to extend the evaluation to skilled nursing facility populations and external cohorts to assess generalizability. Second, the current implementation relies on complete clinical feature availability, which may limit its applicability in real‑world settings where clinical measurements are often incomplete. In future work, we intend to extend \ours to accommodate missing data, for example by incorporating native missing‑value handling within the foundation model.

%\noindent \textbf{Ethical Considerations.} % No need in this version.
%\clearpage
%\input{sections/limitations}

% acknowledgement
%\section*{Acknowledgment}
%Research reported in this work was partially supported by the Atlanta Interdisciplinary Artificial Intelligence Network's 2024 Seed Grant. The content is solely the responsibility of the authors and does not necessarily represent the official views of the the Atlanta Interdisciplinary Artificial Intelligence Network.

% Custom bibliography entries
\bibliographystyle{IEEEtran}
\bibliography{custom}

% appendix
%\clearpage
%\appendix
%\appendixpage
%\input{sections/appendix}

\end{document}